\documentclass[universe,article,accept,pdftex,moreauthors]{Definitions/mdpi} 
\usepackage{longtable}

\usepackage{subfigure} 
\makeatletter
\renewcommand{\@thesubfigure}{\normalsize(\textbf{\alph{subfigure}})}
\makeatother
\firstpage{1} 
\pubvolume{1}
\issuenum{1}
\articlenumber{0}
\pubyear{2023}
\copyrightyear{2023}
 \externaleditor{Academic Editor(s): Name}
\datereceived{30 October 2023 } 
\daterevised{9 December 2023 } 
\dateaccepted{17 December 2023 } 
\datepublished{ } 
\hreflink{https://doi.org/} 

\Title{Mass Distribution and Maximum Mass of Neutron Stars: Effects of Orbital Inclination Angle}

\TitleCitation{Mass Distribution and Maximum Mass of Neutron Stars: Effects of Orbital Inclination Angle}

\Author{Lívia S. Rocha *$^{,\dagger}$\orcidA, Jorge E. Horvath
 \orcidB, Lucas M. de S\'a \orcidC, Gustavo Y.
Chinen \orcidD, \mbox{Lucas G. Bar\~ao \orcidE} \mbox{and Marcio G. B. de Avellar} \orcidF}

\AuthorNames{Lívia S. Rocha, Jorge E. Horvath, Lucas M. de S\'a, Gustavo Y. Chinen, Lucas G. Bar\~ao and Marcio G.B. de Avellar}

\AuthorCitation{Rocha, L.S.; Horvath, J.E.; \mbox{de S\'a,} L.M.; Chinen, G.Y.; Bar\~ao, L.G.; \mbox{de Avellar, }M.G.B.}

\address[1]{Departamento de Astronomia, Instituto de Astronomia, Geofísica e Ciências Atmosféricas (IAG), Universidade de S\~ao Paulo,
R. do Mat\~ao 1226, C. Universit\'aria,  S\~ao Paulo 05508-090, Brazil; \\
foton@iag.usp.br (J.E.H.); lucasmdesa@usp.br (L.M.d.S.); gutavo.chinen@usp.br (G.Y.C.); \mbox{lucasgbarao@usp.br (L.G.B.); } marcio.avellar@alumni.usp.br (M.G.B.d.A.)
}

\corres{\hangafter=1 \hangindent=1.0em \hspace{-1em} Correspondence: livia.silva.rocha@usp.br}

\abstract{Matter at ultra-high densities finds a physical realization inside neutron stars. One key property is their maximum mass, which has far-reaching implications for astrophysics and the equation of state of ultra dense matter. In this work, we employ Bayesian analysis to scrutinize the mass distribution and maximum mass threshold of galactic neutron stars. We compare two distinct models to assess the impact of assuming a uniform distribution for the most important quantity, the cosine of orbital inclination angles ($i$), which has been a common practice in previous analyses. This prevailing assumption yields a maximum mass of $2.25$~$M_\odot$ (2.15--3.32~$M_\odot$ within $90\%$ confidence), with a strong peak around the maximum value. However, in the second model, which indirectly includes observational constraints of $i$, the analysis supports a mass limit of $2.56^{+0.87}_{-0.58}~M_\odot$ ($2\sigma$ uncertainty), a result that points in the same direction as some recent results gathered from gravitational wave observations, although their statistics are still limited. This work stresses the importance of an accurate treatment of orbital inclination angles, and contributes to the ongoing debate about the maximum neutron star mass, further emphasizing the critical role of uncertainties in the individual neutron star mass determinations.}


\begin{document}


\section{Introduction}\label{sec:intro}
More than fifty years after the detection of the first pulsar \cite{Hewish}, these compact objects continue to challenge our understanding. The study of neutron stars (hereafter NSs), of which pulsars are a subset, provides insights into fundamental aspects of the Universe, ranging from the state of matter at ultra-high densities to interstellar enrichment with heavy elements. Additionally, they offer the most precise tests for General Relativity \mbox{(GR) \cite{Kramer2021}} and contribute to the understanding of stellar evolution and binary interactions. The importance of studying such stars is clear, and the last 10--15 years of research have brought about paradigm shifts, correcting previously unquestioned beliefs in the field.

The proposal that NSs are formed at the end of the life of massive progenitor stars, initially suggested by Baade and Zwicky \cite{Baade&zwicky}, coupled with their known associations with supernova remnants \cite{cameron1959}, helped construct an evolutionary scenario for their formation. Despite an initial variety of ideas \cite{nomoto1984evolution}, the prevailing notion was that NSs are formed from the collapse of an Iron ($^{56}$Fe and neighbour isotopes) core, developed at the center of progenitor stars with initial masses ranging from 8 to 25 $M_\odot$. A misleading association of the iron core {before the collapse} with the Chandrasekhar limit, together with early mass measurements, contributed to establishing a paradigm of a unique formation channel for NSs, with a mass scale around $1.4~M_\odot$ and minimal dispersion \cite{finn1994}.

Simultaneously, the existence of a mass limit predicted by the Tolman--Oppenheimer--Volkoff (TOV) equation, derived from the theory of GR and not from fermion degeneracy, sparked extensive discussions. Since the determination of this limit relies on the equation of state (EoS) describing matter at extreme conditions, and this EoS cannot be securely determined from first principles or terrestrial experiments, an accurate maximum mass could not be determined. In 1974, \citet{rhoadesandruffini1974} proposed an ``absolute'' limit under the assumption of a static and isotropic metric. Assuming the equation of state above a certain density to be the \textit{stiffest} possible\endnote{The equation of state needs to obey causality. The stiffest EoS is the one in which the sound speed in the medium equals the speed of light.}, they established an upper threshold for NS masses at $3.2~M_\odot$, which has since been adopted to distinguish NSs from black holes (BHs). While uncertainties in early X-ray mass measurements did not forbid the existence of high masses, and some theoretical studies suggested the possibility of a few physical EoSs leading to a maximum mass around $2~M_\odot$, it became a consensus in the scientific community that, for evolutionary reasons, NS masses should not exceed the ``canonical'' value of $1.4~M_\odot$, in agreement with the first precise mass measurements \cite{finn1994, thorsett1999neutron}. Or, in other words, that the $1.4~M_\odot$ was the value imprinted at birth by collapse physics.

However, observational efforts have steadily increased the number of measured masses over the years. For over a decade now, it has been known that the mass range covered by NSs is much larger than previously expected, with the current interval spanning from $1.17~M_\odot$ to values exceeding $2.0~M_\odot$, a much broader range than was previously thought possible. The first pulsar discovered with a mass that deviates significantly from the ``typical'' value was Vela X-1, with $1.86 \pm 0.16~M_\odot$ \cite{Barziv2001}. In the following years, a few other potentially massive NSs started being discovered, as is the case of PSR J0751+1807 ($m=2.1\pm0.2~M_\odot$) \cite{nice2005}, PSR B1516+02B ($m = 2.08 \pm 0.19~M_\odot$) \cite{freire2008a}, PSR J1748-2021B ($m = 2.74 \pm 0.21~M_\odot$) \cite{freire2008b}, and PSR J1614-2230, with $m = 1.97 \pm 0.04~M_\odot$ \cite{demorest2010}, which was considered the most accurate inference among the massive ones at that time. The discovery of PSR J0740+6620 \cite{cromartie2019relativistic}, with $m = 2.14_{-0.09}^{+0.10}~M_\odot$, proved once and for all the existence of NS masses >$2.0~M_\odot$, highlighting the issue of the maximum mass.

From an evolutionary standpoint, the presence of a broad mass range was soon associated with different formation mechanisms. An analysis by \citet{schwab2010} with a selected sample of well-constrained masses (uncertainties less than $\lesssim$0.025~$M_\odot$) revealed a double-peaked distribution, where a group of NSs centered at $\sim$1.35 $M_\odot$ was associated with the ``standard'' Fe {\it core-collapse supernova}, while the second group clustered around $\sim$1.25 $M_\odot$, linked to the {\it electron-capture supernova} scenario, was expected to occur in degenerate cores of $O-Ne-Mg$ \citep{nomoto1984evolution, podsiadlowski2004effetcs}. Although subsequent analyses with the complete sample of NS masses did not detected this lower peak, it is highly probable that it occurs for progenitor stars with initial masses between 8 and 10 $M_\odot$ \cite{hiramatsu2021}, exploding {\it via} electron capture onto a degenerate $O-Ne-Mg$ core. Instead, these analyses found a second peak around 1.75--1.80 ~$M_\odot$ \cite{zhang2011study, valentim2011mass, kiziltan2013neutron, ozel2016masses}, in addition to the dominant one, at $\sim$1.4 $M_{\odot}$. This massive group was rapidly associated with accretion onto NSs in binary systems, as expected from the existence of millisecond pulsars and other systems containing an NS. Even though accretion is still a prime candidate for the origin of massive NSs, recent works have also demonstrated the possibility of forming massive NSs directly from heavier iron cores \cite{burrows2021}, while others have shown that smaller Fe cores can collapse, leaving behind a 1.17~$M_\odot$ NS \cite{suwa2018}.

One particular class of binary systems, the ``spiders'', are prime candidates to populate the high-mass interval. ``Spiders'' are close binary systems ($P_b < 1$ day), where a millisecond pulsar orbits a low-mass companion star that is in the process of having its envelope ablated away by the pulsar's wind\endnote{This is the reason why these systems are called {\it spiders}, in an analogy with the black widow and redback spiders, which are known to kill and devour their male partners.}. If the donor companion has a mass of 0.1--0.5~$M_\odot$, it is classified as a \textit{redback}, while those with companion masses $\lesssim 0.05~M_\odot$---and where the accretion has stopped---are named \textit{black widows}. These systems are expected to experience a large accretion phase in the early-stage phases, accumulating a great amount (>0.8~$M_\odot$) of mass in the most extreme cases \cite{kandel2022optical}. Evidence suggests that these systems can host the most massive NSs in the Universe \cite{Linares, horvath2020spiders}. The most recent massive spider discovered is PSR J0952-0607, with $m = 2.35\pm0.17~M_\odot$ \cite{Romani2022heaviest}, placing the maximum NS mass as $>$2.19~$M_\odot$ ($1\sigma$ confidence).

Finally, NSs may also be formed through the (single-degenerate) \textit{accretion-induced collapse} (AIC) scenario, where a massive white dwarf (WD) exceeds its mass limit and collapses without igniting carbon, or the double-degenerate AIC, where two WDs merge. Although these events have never been positively identified, population synthesis has yielded the expectation of approximately $10^{7}$ pulsars formed by AIC in the single-degenerate channel and a few times this figure comes from the double-degenerate channel \citep{DongDong}. Additionally, the long gamma-ray burst GRB 211211A fed the possibility of an NS-WD merger \cite{zhong2023grb}, leaving behind a magnetar, another feasible scenario to populate the second peak of masses \mbox{around $\sim$1.8 $M_{\odot}$.}

The detection of gravitational waves (GW) from the merger of two NSs in 2017, GW170817 \cite{abbott2017}, opened a new era in multimessenger and multiwavelength astronomy, providing a new tool to set constraints on NS physics. Together with galactic determinations, constraints placed by GW170817 contributed to placing the limit of masses below \mbox{2.2--2.3$ ~M_\odot$ \cite{margalit2017, alsing2018evidence, shibata2019, shao2020a, shao2020b}}. Later on, the detection of GW190814 \cite{abbott2020gw190814}, where one of the components that merged with a $23~M_\odot$ BH had a mass of $2.59_{-0.09}^{+0.08}~M_\odot$, raised a tension about whether the threshold of the maximum mass should be above or below 2.2--2.3 ~$M_\odot$. The work of \citet{nathanail2021} concluded that the secondary of GW190814 needs to be a BH in order to avoid contradictions with the post-merger observations of GW170817. At the same time, an analysis made by the Ligo--Virgo (LV) Collaboration \citep{abbott2021population} of the binary BH (BBH) merging population in the second catalog found GW190814 to be an outlier, {i.e.}, to be likely originated from an NS--BH merger. In addition, from an analysis of the available LV NS--BH events---and assuming GW190814 to be one of them--the maximum mass of a non-spinning NS was found to be $2.7^{+0.5}_{-0.4}~M_{\odot}$ \citep{ye2022inferring}. A recent result combined constraints from all possible astronomical approaches, and derived a maximum mass in the range of 2.49--2.52~$M_\odot$ \cite{Ai2023constraints}, providing independent evidence supporting the existence of extremely heavy NS masses.

In this article, we focus on analyzing the up-to-date sample of NS masses found in binary systems in the Galaxy and in Globular Clusters, and the impact the orbital inclination angle has on the conclusions derived from this kind of analysis, especially in the estimation of the maximum mass ($m_{max}$). We start in Section \ref{sec:mass_meas} with an overview of all available measurement methods and a discussion of the detailed features leading to individual masses and uncertainties. In Section \ref{sec:bayesian_analysis}, we describe the statistical method and the ``accuracy-dependent'' and ``accuracy-independent'' models used to analyze the whole distribution, now featuring 125 members (listed in Appendix \ref{app:table}). In Section \ref{sec:results}, we present our results and compare the posterior distributions of each model. Finally, we draw some general conclusions in the last section.
It is important to emphasize the difference between the TOV mass ($M_{TOV}$), supported by non-rotating pulsars, and the $m_{max}$, which is derived from our analysis and is the maximum supported by pulsars with maximum uniform rotation\endnote{The difference is irrelevant for the whole sample of 3000+ pulsars
known today, which would need to rotate much faster to hold an excess of
mass over the $M_{TOV}$}. Furthermore, according to the \textit{quasi-universal} relation derived in \citet{Most2020}, $m_{max}$ can exceed by a factor of up to 1.2 the $M_{TOV}$ value.

\section{Neutron Star Mass Measurements}\label{sec:mass_meas}

The majority of NSs are observed as pulsars, rapidly-rotating and highly magnetized NSs emitting beams of radiation along their magnetic axes. This emission is observed on Earth as a pulse due to the lighthouse effect. The remarkable regularity of these pulses, with pulsars being one of the most stable clocks in the observable universe, makes {\it pulsar timing} the most accurate method for determining their masses and testing fundamental physics. This procedure involves monitoring the times of arrival (ToAs) of pulses over an extended period of time to determine the pulsar's rotation period. Thanks to the regularity, small deviations in ToAs are detectable with precision and are indicative of the presence of a companion. The greater the number of collected ToAs, the greater will be the precision achieved. Hence, several years of observations are necessary to achieve a high \mbox{precision \cite{lorimer2005}.}

Currently, more than 3300 radio pulsars have been identified (see an online catalogue at ATNF \cite{atnf}), yet only a few of their features can be directly inferred from observations, and mass measurements are possible for only a small fraction of the total sample. The \textit{Neutron star Interior Composition ExploreR} (NICER), a telescope placed on board the International Space Station in 2017, facilitates timing- and rotation-resolved spectroscopy of thermal and non-thermal X-ray emissions from NSs. Recently, NICER enabled precise measurements of radii and masses for two pulsars, namely PSR J0740+6620 and PSR J0030+0451 \citep{riley2019nicer, 2021ApJ...918L..27R}. Despite this being a promising method, especially for isolated pulsars, the dominant means for inferring NS masses continues to be the study of orbital motion in binary systems determined through pulsar timing of radio sources.

The ToAs reveal the orbital properties of the system expressed in terms of Keplerian parameters: orbital period ($P_B$), eccentricity ($e$), semi-major axis projection onto the line of sight ($x = a\sin{i}$), and the time ($T_0$) and longitude ($\omega$) of periastron. From Kepler's third law, a mass function containing the pulsar mass ($m_p$), the companion mass ($m_c$), and the inclination angle ($i$) of the system can be written down as:
\begin{equation}
    f_p = \frac{(m_c\sin{i})^3}{(m_p+m_c)^2} = \frac{4\pi^2}{T_\odot}\frac{x^3}{P_B^2},
\end{equation}
where $T_{\odot} \equiv G M_{\odot}/c^3 = 4.925490947~\upmu$s ($G$ is the gravitational constant, $c$ is the speed of light, and $M_{\odot}$ is the solar mass). If the mass function of the pulsar and the companion are measured ($f_p, f_c$), along with the mass ratio ($q = m_p/m_c$), the individual masses of the system can be determined, provided the inclination angle $i$ is known. However, until now we were only able to set precise constraints for both masses in a few particular cases, because of the difficult estimation of the inclination angle $i$, which we now address.

\subsection{Orbital Inclination Angle}\label{sec:orbital_incl}

Since binary stars have small angular sizes, it is challenging to fully resolve the orbit. Consequently, if the radial and transverse velocities of the system cannot be measured, the orbital inclination of the orbit with respect to the plane of the sky ($i$) cannot be \mbox{directly determined.}

However, these systems often display variability of the companion light curves, from which an inclination-dependent radial velocity can be estimated. Pulsar mass estimates via radial velocity measurements depend on the inclination angle $i$ of the orbit with respect to the line of sight, through the relation:
\begin{equation}
    m_p \propto \frac{1}{\sin^3 i}.
\end{equation}

Therefore, a high uncertainty in the inclination angle will result in a high uncertainty in the mass measurement. Although it is difficult to precisely determine $i$, constraints can be set if, for example, an eclipse is observed through spectroscopy of the companion. 

\subsection{Relativistic Binaries}\label{sec:relativistic_binary}
In the special case of compact binaries, where the companion is a WD or an NS, relativistic effects may influence the orbit and can sometimes be measured. These effects are described in terms of the so-called post-Keplerian (pK) parameters, defined as:
\begin{enumerate}
    \item Orbital period decay, $\Dot{P}_b$:
        \begin{equation}
        \Dot{P}_b = -\frac{192\pi}{5} \left( \frac{P_b}{2\pi T_{\odot}}\right)^{-\frac{5}{3}} \left( 1 + \frac{73}{24}e^2 + \frac{37}{96}e^4 \right) \left(1 - e^2 \right)^{-\frac{7}{2}} \frac{m_p m_c}{m^{1/3}};
        \label{eq1.2}
        \end{equation}
    \item Range of Shapiro delay, $r$:
        \begin{equation}
        r = T_{\odot}m_c;
        \label{eq1.3}
        \end{equation}
    \item Shape of Shapiro delay, $s$:
        \begin{equation}
        s = \sin{i} = x_p \left(\frac{P_b}{2\pi} \right)^{-2/3} \frac{m^{2/3}}{T_{\odot}^{1/3}m_c};
        \label{eq1.4}
        \end{equation}
    \item ``Einstein delay'', $\gamma$:
        \begin{equation}
        \gamma = e \left( \frac{P_b}{2\pi} \right)^{1/3} T_{\odot}^{2/3}~ \frac{m_c \left(m_p + 2m_c\right)}{m^{4/3}};
        \label{eq1.5}
        \end{equation}
    \item Advance of periastron, $\Dot{\omega}$:
        \begin{equation}
        \Dot{\omega} = 3\left(\frac{P_b}{2\pi} \right)^{-5/3} \left( 1 - e^2 \right)^{-1} \left( m~ T_{\odot} \right)^{2/3}.
        \label{eq1.6}
        \end{equation}
\end{enumerate}

If at least two pK parameters are measured, the component masses can be individually determined. When more pK parameters are measured, it is also possible to test GR with very high precision, as demonstrated in \citet{Kramer2021}. Accretion torques in binary systems can circularize the orbits, resulting in many NS binaries with extremely low eccentricities, hampering the measurement of $\Dot{\omega}$ and $\gamma$. On the other hand, the Shapiro delay of the pulses, caused by the gravitational field of the companion, depends on the orbital inclination and is typically relevant for systems with high inclinations. Lastly, orbital decay due to gravitational wave radiation ($\Dot{P_b}$) is only measurable for very tight orbits. All of these conditions make pulsar mass measurements a challenging task. If relativistic effects are too small, they can go undetected even after years of pulsar timing.

\subsubsection*{Shapiro Delay}\label{sec:shapiro}
The Shapiro delay is the increase in light travel time through the curved space--time near a massive body. In binary pulsar systems that have highly inclined (nearly edge-on) orbits, excess delay in the pulse ToA can be observed when the pulsar is situated almost behind the companion during orbital conjunction. In combination with the mass function, the Shapiro delay offers one of the most precise methods with which to directly infer the mass of NSs. 

\subsection{Optical Spectroscopy}\label{sec:optical_sepc}
On the other hand, when the pulsar has an optically bright low-mass companion, such as a Main Sequence or post-Main Sequence star or a WD, phase-resolved spectroscopy of the companion can yield the orbital radial velocity amplitude ($K_c$). When combined with $x$ and $P_b$, this provides the binary mass ratio $q = (m_p/m_c) = (K_c/K_p)$. In the case of WD companions, their radii can be estimated when the distance ($d$) to Earth is known, along with the optical flux ($F_O$) and effective temperature ($T_{eff}$) measurements, as:
\begin{equation}
    R_{WD} = \left(\frac{F_O}{\sigma}\right)^{1/2}\left(\frac{d}{T^2_{eff}}\right),
\end{equation}
where $\sigma$ is the Stefan--Boltzmann constant. Their masses can thus be estimated by combining the effective temperature with the surface gravity obtained from an atmosphere model, which provides a model-dependent method. {Combining $m_c$ with $q$, the pulsar mass is addressed.}

Neutron star masses in spider systems are typically inferred through spectrophotometric methods. The system is filled with intra-binary material, causing the radio pulsation to be scattered and absorbed \citep{kansabanik2021unraveling}. As a consequence, their optical light curves are sensitive not only to the orbital inclination, but also to the heating models of the companion's surface, which can be challenging to predict. A large systematic error in the inclination angle estimate can result in a significant bias in the mass estimate for this class of pulsars.

\subsection{Gamma-Ray Pulsars}\label{sec:gamma_ray_obs} 
Millisecond pulsars also present gamma-ray pulsations \citep{atwood2009large}. In contrast to other wavelengths, it seems unlikely that gamma-rays are absorbed in the diffuse intra-binary material of spider systems. Consequently, the observed gamma-ray eclipses are potentially associated only with the occultation of the pulsar by the companion, providing a more robust determination of the inclination angle. The work of \citet{clark2023neutron} conducted a search for gamma-ray eclipses in 49 confirmed and candidate spider systems, with significant detections in five of them, from which mass determinations were obtained \mbox{(Table \ref{tab:gamma_rays}}). In only one of these five systems, the inclination angle was found to be inconsistent with previous optical modeling. 

\textls[-25]{For PSR B1957+20, photometric observations provided an estimate of $63^{\circ} \lesssim i \lesssim 67^{\circ}$,} resulting in a best fit with a high mass of $m = 2.4 \pm 0.1~M_\odot$ \citep{van2011evidence}. Observations in gamma-rays, on the other hand, require an inclination angle $i > 84.1^{\circ}$, corresponding to a significantly lower mass of $m = 1.81 \pm 0.07~M_\odot$. If this discrepancy is found to be consistent for the most massive objects measured through X-ray/Optical modeling, as well as confirmed for this particular system, it could have a substantial impact on the determination of the maximum mass of NSs and, consequently, on our understanding of matter at ultra-high densities. 

However, it is essential to note that the results from \citet{clark2023neutron} assume that gamma-ray photons detected at Fermi-LAT energies are {\it not absorbed} by the diffuse material and that the wind is isotropic (a questionable simplified assumption). Further studies on intra-binary shocks in spider systems are still necessary to firmly constrain their geometry and confirm such results.

\begin{table}[H]
     \caption{Mass estimates (1$\sigma$ uncertainty) for spider systems with detected eclipses in gamma-rays, derived by \citet{clark2023neutron}. }
    \label{tab:gamma_rays}
    \tabcolsep=2.55cm
    \begin{tabular}{l c}
        \toprule
         \textbf{Pulsar} & \boldmath{$m_p~(M_\odot)$} \\
             \midrule
         B1957+20 & 1.81  $\pm$  0.07\\
        J1048+2339 & 1.58  $\pm$  0.07\\
        J1555$-$2908 & 1.65  $\pm$  0.04\\
        J1816+4510 & 1.90  $\pm$  0.13\\
        J2129$-$0429 & 1.70  $\pm$  0.11\\
      \bottomrule
    \end{tabular}

\end{table}

\section{Analysis of the Mass Distribution}\label{sec:bayesian_analysis}
As mentioned in the Introduction, studying the mass distribution of NSs can provide valuable insights into the evolutionary mechanisms leading to their formation and help constrain their maximum masses. The very first of these analyses was conducted on a small sample of eight NS masses from four double neutron star (DNS) systems, and employed Bayesian inference \cite{finn1994}. Those results indicated that NS masses should predominantly fall within the range of $1.3 < m/M_\odot < 1.6$. Subsequent analyses with a larger sample of 19 NS masses arrived at a similar conclusion \cite{thorsett1999neutron}, with no evidence of a significant dispersion around the mean value of $1.35 \pm 0.04~M_\odot$.

As the search for NSs continued and surveys yielded an increasing number of discovered objects, the landscape began to change. Over the years, different research groups employed frequentist and Bayesian inference techniques to extract information from the mass distribution of NSs \cite{valentim2011mass, zhang2011study, kiziltan2013neutron, ozel2016masses, antoniadis2016, alsing2018evidence, farrow2019, shao2020a}. All these analyses consistently revealed the presence of a double-peaked distribution. These findings directly challenged the old idea of a single formation channel, since such a scenario could hardly account for the entire range of observed masses, suggesting that different processes are at work.

In this work, we employ a Bayesian analysis of an updated sample of NS masses using the advanced technique of \textit{Markov Chain Monte Carlo} (MCMC) simulations. This technique allows us to determine the \textit{posterior} distribution of a set of unknown parameters ($\boldsymbol{\theta}$) based 
 on the \textit{a priori} information available for each of these parameters, combined with information from observed data. According to Bayes' theorem, the posterior distribution is expressed as $P(\boldsymbol{\theta} | d) = P(\boldsymbol{\theta}) {\cal L}(d | \boldsymbol{\theta})$, where $P(\boldsymbol{\theta})$ represents the \textit{a priori} distribution, and ${\cal L}(d | \boldsymbol{\theta})$ denotes the \textit{likelihood} of the model. Assuming a parameterized model, the \textit{likelihood} can be defined as ${\cal L}(d | \boldsymbol{\theta}) = P(\mathbf{d}^i | m_p^i) P(m_p^i | \boldsymbol{\theta})$, where $\mathbf{d^i}$ is the data. Consequently, the posterior distribution marginalized over pulsar masses ($m_p^i$) is given by:
\begin{equation}\label{eq:trunc_posterior_distr}
    P(\boldsymbol{\theta} | \mathbf{d}) \propto P(\boldsymbol{\theta}) \prod_{i=1}^{N} \int P(\mathbf{d}^i | m_p^i) P(m_p^i | \boldsymbol{\theta})~dm_p^i.
\end{equation}

To account for the double-peaked distribution, recent works have adopted a Gaussian mixture model for the model likelihood, $P(m_p^i | \boldsymbol{\theta})$, with $n$ components. This model has been compared with other distribution families but, so far, no evidence has emerged to reject the preference for a Gaussian mixture model. Furthermore, if we wish to introduce a free parameter to estimate a cutoff in the distribution (motivated by the expectation of a maximum mass), we can employ a {\it truncated} Gaussian mixture model:
\begin{equation}\label{eq:trunc_likelihood}
    P(m_p^i | \boldsymbol{\theta}) = \sum_{j=1}^n r_j \frac{{\cal N}(m_p^i | \mu_j, \sigma_j)}{ \int^{m_{max}}_{m_{min}}{\cal N}(x | \mu, \sigma)dx},
\end{equation}
with $\sum_j^n r_j = 1$ to ensure normalization. The set of model parameters we aim to infer includes the mean ($\mu_j$) and standard deviation ($\sigma_j$) of each Gaussian component, along with their respective amplitudes ($r_j$) and the maximum mass ($m_{max}$) set as a free parameter, $\boldsymbol{\theta} = \{\mu_j, \sigma_j, r_j, m_{max}\}$, with $j = \{1, n\}$. In the following subsections, we will delve into the results of our analysis assuming two different individual pulsar distributions, $P(\mathbf{d}^i | m_p^i)$. To sample our posterior distributions, we employ an MCMC algorithm with \citet{sdstan}.

\subsection{``Accuracy-Dependent'' Model}\label{sec:alsing_model}
As discussed in Section \ref{sec:mass_meas}, the most accurately determined pulsar masses are those where relativistic effects are revealed, and two or more pK parameters are observed. However, in general, satisfying these requirements is not straightforward, and in many instances only mass limits can be established. In this work, we employed the models used in the works of \citet{antoniadis2016, alsing2018evidence, shao2020a}, {which we refer to as \textit{accuracy-dependent} models}, where the pulsar mass likelihood depends on whether or not two or more pK parameters are measured, as we describe below.

For systems where the pulsar mass is accurately determined from observations, we assume the likelihood to be a normal distribution, ${\cal N}(m_i, \sigma_{m_i})$, with mean and standard-deviation values provided in the sixth column of Table \ref{tab:mass}. In cases where, despite the mass function $f$, only the $\Dot{\omega}$ is constrained, it is possible to determine the total mass ($m_t$) of the system, given in the fourth column of the same table. In such cases, the pulsar's mass is marginalized as
\begin{equation*}
    \begin{split}
    P(d|m_p) &\propto \int \int P(\hat{m}_t, \hat{f}|m_p, m_t, i)P(m_t)P(i)di~dm_t \\
            &\propto \int \int P(\hat{m_t}|m_t)P(\hat{f}|f(m_p, m_t, i))P(m_t)P(i)di~dm_t  \\
            &\propto \int \int exp\left(-\frac{(m_t - \hat{m}_t)^2}{2\sigma_{m_t}^2} \right) \delta(f(m_p, m_t, i) - \hat{f}) \sin i~di~dm_t,  \\
    \end{split}
\end{equation*}
where $\hat{m}_t$, $\sigma_{m_t}$, and $\hat{f}$ are the measured total mass, its uncertainty, and the measured mass function, respectively. We assume that the total mass and the mass function are independent, and that $m_t$ has Gaussian uncertainties. Integrating the last line of the above equation over $i$ with a flat \textit{a priori}, the individual pulsar mass can be marginalized from:
\begin{equation}\label{eq:total_mass_like}
    P(d|m_p)= \int exp\left( -\frac{(m_t - \hat{m}_t)^2}{2\sigma_{m_t}^2}\right) \frac{m_t^{4/3}}{3(m_t - m_p)^2f^{1/3}\sqrt{1-\frac{f^{2/3}m_t^{4/3}}{(m_t - m_p)^2}}}dm_t.
\end{equation}
Note that the above equation is slightly different from Equation (3) in \citet{alsing2018evidence}, as suggested by \citet{farr2020population}.

Finally, if only the mass ratio ($q$) can be determined through phase-resolved optical spectroscopy, in addition to $f$, the likelihood of pulsar mass is given as:
\begin{equation*}
    \begin{split}
    P(d|m_p) &\propto \int \int P(\hat{q}, \hat{f}|m_p, q, i)di~dq \\
             &\propto \int \int P(\hat{q}|q)P(\hat{f}|f(m_p, q, i))P(q)P(i)di~dq \\
             &\propto \int \int exp\left(- \frac{(q-\hat{q})^2}{2\sigma_q^2}\right)\delta (f(m_p, q, i) - \hat{f}) \sin i~di~dq,
    \end{split}
\end{equation*}
where $\hat{q}$ and $\sigma_q$ are the measured mass ratio and its uncertainty. Here, we assume that the mass ratio and the mass function are independent and that $q$ has Gaussian uncertainties. Integrating the last line over $i$ and assuming again a flat \textit{a priori}, we are led to:
\begin{equation}\label{eq:mass_ratio_like}
    P(d|m_p) = \int exp\left(-\frac{(q-\mu_q)^2}{2\sigma_q^2} \right)\frac{(1+q)^{4/3}}{3f^{1/3}m_p^{2/3}q^2\sqrt{1-\left(\frac{f}{m_p}\right)^{2/3}\frac{(1+q)^{4/3}}{q^2}}}dq.
\end{equation}

The key assumption for deriving Equations \eqref{eq:total_mass_like} and \eqref{eq:mass_ratio_like} is that $\cos i$ follows an isotropic distribution, which means it is subject to a uniform \textit{a priori} assumption, allowing $i$ to span any value between $0^\circ$ and $90^\circ$.

\subsection{``Accuracy-Independent'' Model}\label{sec:gaussian_model}
Given the requirement for a uniform distribution for $\cos i$ in the previous model, and considering the substantial impact that the orbital inclination angle has on the pulsar mass estimation (as discussed in Section \ref{sec:orbital_incl} and shown later in Section \ref{sec:results}), we looked for an alternative model that could avoid such an assumption. In the ``accuracy-independent'' model, we assume all pulsar masses listed in Table \ref{tab:mass} to be modeled with a normal distribution, as performed in previous works \cite{valentim2011mass, horvath2020book,  livia2023thesis}. To illustrate the procedure, observations of PSR 2S 0921-630 {(the first in our table)}, for example, found it to have a mass of $1.44\pm0.10~M_\odot$ \cite{steeghs2007mass}. We then modeled this system with a normal distribution, with a mean value of $1.44$ and a standard-deviation of 0.1, ${\cal N}(1.44, 0.10)$, and so on. Systems without any  constraints on the individual mass were naturally excluded from this analysis. 

Although to model all masses with Gaussians may appear less robust, it is worth noting that the pulsar mass values reported in Table \ref{tab:mass} were calculated while taking into account all observational measurements and constraints of Keplerian and post-Keplerian parameters, which are tightly tied, available for each particular system, including $i$. Furthermore, Bayesian statistics is known to weight the sampling from data uncertainty, {i.e.}, data with large uncertainties will have a lower weight in posterior distributions than those with small uncertainties. {In this sense, mass determinations derived from spectrophotometry will have a lower weight in the posterior results than those derived from Shapiro delay, \mbox{for example.}}

\section{Results}\label{sec:results}
The summaries of marginalized distributions for the ``accuracy-dependent'' and ``accuracy-independent'' models are presented in Tables \ref{tab:alsingsampling} and \ref{tab:gaussiansampling}, respectively. With the exception of $m_{\max}$ in the last row, for which we report the mode value, the second column displays the mean value for each parameter, followed by its respective standard deviation. The third column shows the ``highest posterior density interval'' (HPDI) with $94\%$ probability, indicating the shortest range of values that encompasses the given probability.

\begin{table}[H]
\caption{Summary of marginal posterior distribution for the ``accuracy-dependent'' model. With the exception of $m_{max}$, the second column displays the mean value, followed by respective standard deviation and the $94\%$ highest posterior density interval, which defines the lowest interval that comprises $94\%$ of the probability. For $m_{max}$, the value shown in column 2 is the mode.} 
\label{tab:alsingsampling} 
\tabcolsep=1.2cm                 
\begin{tabular}{l c c c c c } 
\toprule         
      & \textbf{Mean} & \textbf{SD} & \boldmath{$94\%$} \textbf{HPDI} \\
\midrule  
    $r_1$ & $0.576$ & $0.096$ & $0.398$--$0.759$ \\
    $r_2$ & $0.424$ & $0.096$ & $0.241$--$0.602$ \\
    $\mu_1$ & $1.354$ & $0.025$ & $1.310$--$1.402$ \\
    $\mu_2$ & $1.830$ & $0.149$ & $1.604$--$2.137$ \\
    $\sigma_1$ & $0.088$ & $0.021$ & $0.051$--$0.129$ \\
    $\sigma_2$ & $0.291$ & $0.082$ & $0.165$--$0.470$ \\
    $m_{max}$ & $2.25$ & $0.15$ & $2.122$--$3.246$ \\
\bottomrule  
\end{tabular}
\end{table}
\vspace{-8pt}

\begin{table}[H]
\caption{Summary of marginal \textit{posterior} distribution for the ``accuracy-independent'' model. With the exception of $m_{max}$, the second column displays the mean value, followed by the respective standard deviation and the $94\%$ highest posterior density interval, which defines the lowest interval that comprises $94\%$ of the probability. For $m_{max}$, the value shown at column 2 is the mode.} 
\label{tab:gaussiansampling} 
\tabcolsep=1.2cm                   
\begin{tabular}{l c c c c c } 
\toprule           
      & \textbf{mean} & \textbf{SD} & \boldmath{$94\%$ }\textbf{HPDI }\\
    \midrule
    $r_1$ & $0.539$ & $0.086$ & $0.371$--$0.692$ \\
    $r_2$ & $0.461$ & $0.086$ & $0.308$--$0.629$ \\
    $\mu_1$ & $1.351$ & $0.022$ & $1.308$--$1.392$ \\
    $\mu_2$ & $1.816$ & $0.073$ & $1.677$--$1.954$ \\
    $\sigma_1$ & $0.084$ & $0.019$ & $0.048$--$0.119$ \\
    $\sigma_2$ & $0.260$ & $0.053$ & $0.155$--$0.353$ \\
    $m_{max}$ & $2.56$ & $0.37$ & $1.910$--$3.268$ \\
\bottomrule  
\end{tabular}
\end{table}

As is apparent in both Tables, and in line with previous analyses available in the literature, both samplings yield a bimodal distribution featuring two distinct ``groups'' of NSs. The first group is centered around $\sim$$1.35~M_\odot$ with a small standard deviation of roughly $0.09~M_\odot$. In the second group, the NSs cluster at approximately $1.8\pm0.26~M_\odot$. The predominance of objects in the first group is discernible from the amplitude values $r$. The real divergence between both models becomes apparent in the marginalized distribution of $m_{max}$, which we will explore further below.

\begin{figure}[H]
    \begin{subfigure}
        \centering
        \includegraphics[width=.45\textwidth]{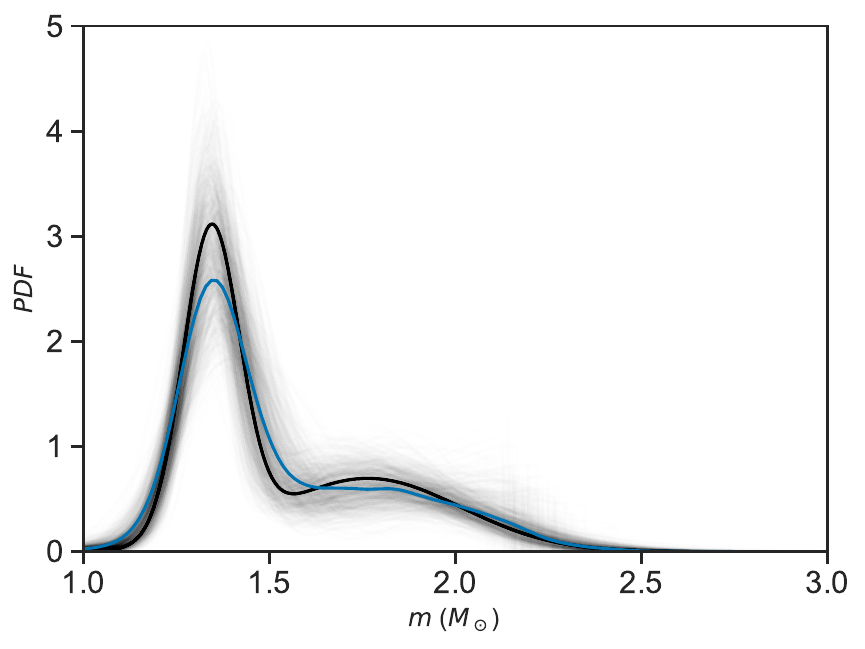}\label{fig:alsingposterior}
    \end{subfigure} %
    \begin{subfigure}
        \centering
        \includegraphics[width=.45\textwidth]{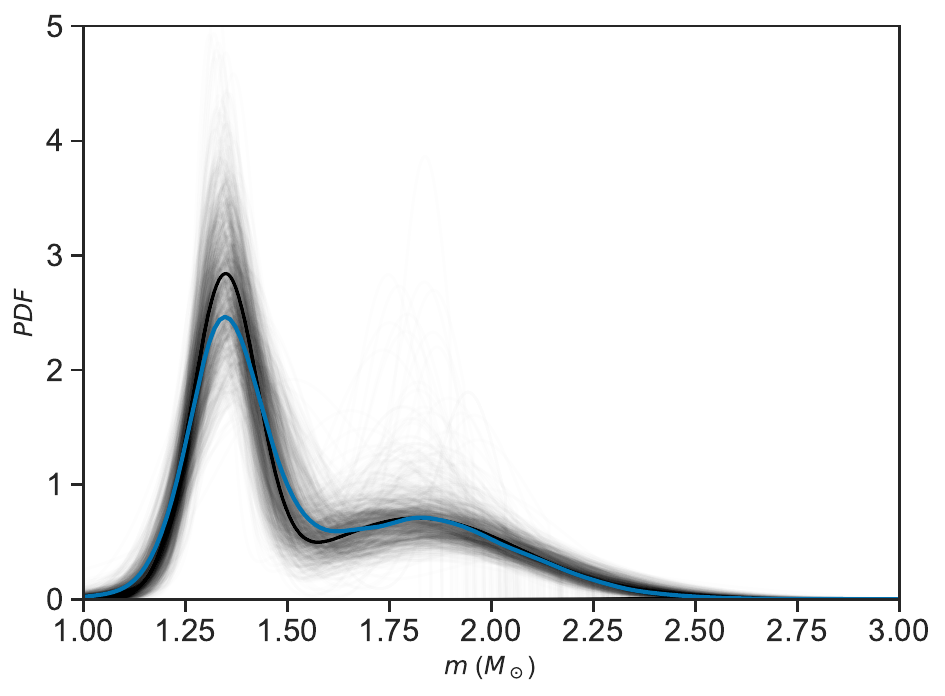}\label{fig:gaussianposterior}
    \end{subfigure} %
\caption{Grey lines represent 1000 posterior samples drawn from the truncated models summarized in Table  \ref{tab:alsingsampling} (left panel) and  Table \ref{tab:gaussiansampling} (right panel). The blue curve is the posterior mean distribution and the black line is the maximum posterior distribution.}\label{fig:posteriordistributions}
\end{figure}

Figure \ref{fig:posteriordistributions} provides a visual representation of the uncertainties in each parameter, summarized in Table  \ref{tab:alsingsampling} (left panel) and  Table \ref{tab:gaussiansampling} (right panel). In the figure, light grey lines depict 1000 posterior samples of the sampled pulsar mass distribution. The \textit{Maximum Posterior Probability}, depicted in black, corresponds to the estimate that aligns with the mode of the \textit{posterior} distribution. The blue line represents the average mass distribution, which is constructed assuming the mean values of each parameter. In contrast to the findings in \citet{alsing2018evidence} and \citet{shao2020a}, our analysis did not reveal a sharp cut-off in the posterior distributions for either of the models.

\subsection{Effects on Individual Masses}\label{sec:eff_ind_mass}
To investigate the differences between the two models defined above, we analyzed the impact of a uniform distribution for $\cos i$ on the sampling of individual pulsar masses. In the ``accuracy-dependent'' model, if only $f$ and $m_t$ or $q$ are known, the individual masses of pulsars are marginalized from Equations (\ref{eq:total_mass_like}) and (\ref{eq:mass_ratio_like}). For the sake of simplicity, we present in the second and third columns of Table \ref{tab:change_like} the $95\%$  HPDI of masses of four selected systems (namely the pulsars  B1957+20, J1311-3430, B1516+02B and J1748-2021B) derived in the ``accuracy-dependent'' model, using Equations (11) and (12). These systems were known for a long time to be potentially extremely massive ($\geq$2.0~$M_\odot$), and are important---although not exclusively---for the inference of $m_{max}$. We now focus our attention on them to illustrate the impact of a flat distribution over $\cos i$ on posterior estimations.

\begin{table}[H]
\caption{Individual pulsar mass of systems listed in this table are sampled from Equations (\ref{eq:total_mass_like}) and (\ref{eq:mass_ratio_like}), as functions of the total mass (second column) or mass ratio (third column). They are particularly interesting since their masses derived from observations are $\geq$2.0~$M_\odot$ (fourth column).} 
\label{tab:change_like}
\tabcolsep=.75cm                 
\begin{tabular}{l c c c } 
\toprule           
     \textbf{Pulsar} & \boldmath{$m_{mt}~ [M_\odot]$} & \boldmath{$m_q~ [M_\odot]$ }& \boldmath{$m_{observations} ~[M_\odot]$}\\
    \midrule
   B1957+20 & - & 1.16--2.14 & $2.40\pm0.12$\\
    J1311-3430 & - & 1.17--2.14 & $2.22\pm0.10$ \\
   B1516+02B & 1.26--2.15 & - & $2.08\pm0.19$ \\
    J1748-2021B & 1.23--2.59 & - & $2.74\pm0.21$  \\
\bottomrule
\end{tabular}
\end{table}

Below, we provide comments on the inferences for each of the four pulsars listed in Table \ref{tab:change_like}. We compare sampled masses with the results obtained from spectrophotometry modeling, summarized in the fourth column. It is worth noting that, for all four pulsars, the masses sampled from the ``accuracy-dependent'' model are inconsistent with values derived from the original observations (last column).

\subsubsection{PSR B1957+20}
A black-widow system. Analysis of the light curve of the companion star yields a radial--velocity amplitude of $K_2 = 353 \pm 4$ km s$^{-1}$. When combined with the pulsar's mass function, this measurement provides a minimum companion mass of $m_{c, min} = 0.022~M_\odot$. The best-fit values for the mass ratio and inclination angle are $q = 69.2 \pm 0.8$ and $i = 65^{\circ} \pm 2^{\circ}$, respectively. These parameters, when combined, result in a best-fit pulsar mass of $m = 2.40 \pm 0.12~M_\odot$ \citep{van2011evidence}, and a lower limit on the pulsar mass at $1.66~M_\odot$. As seen in the third column of Table \ref{tab:change_like}, the ``accuracy-dependent'' model leads us to a marginalized mass between 1.16--2.14~$M_\odot$, inconsistent with spectroscopic determination. {As we commented in Section} \ref{sec:gamma_ray_obs},
{the work in} \cite{clark2023neutron} {places a constraint of $50^{\circ} < i < 85^{\circ}$, resulting in a mean mass at $1.81 \pm 0.07~M_\odot$. Further investigation into the geometry and emissions of spider systems is necessary to pinpoint the correct mass value.}

\subsubsection{PSR J1311-3430}
Until recently, the mass of this black-widow system was determined using a light-curve analysis, resulting in a mass of $m = 2.63^{+0.3}_{-0.2}~M_\odot$ \cite{romani2012psr}. However, constraints on the inclination angle $i$ were poor. More recently, \citet{kandel2022optical} conducted an analysis of heating models for the light curve, leading to improved determinations of the NS masses. In their preferred model, the inclination angle is estimated to be $i = 68.7^\circ \pm 2.1^\circ$. With a radial velocity amplitude of $K_2 = 641.2 \pm 3.6$ and a companion mass of $m_c = 0.012 \pm 0.006$, the pulsar's mass is inferred to be $m = 2.22 \pm 0.10~M_\odot$, {above the $95\% HPDI$ derived in the ``accuracy-dependent'' model}.

\subsubsection{PSR B1516+02B}
This pulsar is located in the globular cluster NGC 5904 and is part of a binary system with a companion that is either a WD or a low-mass {main sequence (MS) star. The binary system has a total mass of $2.29 \pm 0.17~M_\odot$, leading to a best-fit pulsar mass estimate of $m = 2.08 \pm 0.19~M_\odot$ \citep{freire2008a}. There is a $90\%$ probability that the pulsar's mass is greater than $1.82~M_\odot$, and a $0.77\%$ probability that the inclination angle is low enough for the pulsar's mass to fall within the range of 1.20--1.44~$M_\odot$.

\subsubsection{PSR J1748-2021B}
This pulsar is part of a massive binary system with a total mass of $2.92 \pm 0.20~M_\odot$, which was obtained from precise measurements of $\Dot{\omega}$ under the assumption of a fully relativistic system \citep{freire2008b}. The probability that the pulsar's mass falls within the range of 1.20--1.44 $M_\odot$ is only $0.10\%$, requiring an extremely low orbital inclination. The median mass of the companion star is estimated to be 0.142 with lower and upper 1$\sigma$ limits at 0.124 and 0.228 $M_\odot$, respectively. This range of companion masses suggests that the companion star could be a WD or a non-evolved MS star, which implies a pulsar mass of $2.74~M_\odot$.

\subsection{Effects on the Maximum Mass}\label{sec:eff_max_mass}
We now proceed to investigate the impact of a uniform distribution for $\cos i$ on the posterior distribution of $m_{max}$. Following the approach of \citet{alsing2018evidence} and \citet{shao2020a}, we obtained the marginalized posterior distribution shown by the solid line in the left panel of Figure \ref{fig:maximum_masses}, with a maximum of $\sim$2.2~$M_\odot$. On the other hand, the marginalized posterior distribution for the model where all pulsars are assumed to be normally distributed (``accuracy-independent'' model) is shown by the solid line in the right panel of Figure \ref{fig:maximum_masses}, with a maximum of around $2.6~M_\odot$. In addition to the maximum values that differ considerably between the two treatments, the shape of the $m_{max}$ distribution changes significantly too, as we can see in Figure \ref{fig:maximum_masses}. 

For both models, we verified the impact of changing the masses of PSR B1957+20, PSR J1048+2339, PSR J1555-2908, and PSR J2129-0429 by the values derived from $\gamma$-ray observations, listed in Table \ref{tab:gamma_rays}. Since in the analysis of \citet{clark2023neutron} the only value that changed considerably compared to previous estimates was that of PSR B1957+20, we did not expect a significant change in the overall results, as confirmed by the dashed line in both panels of Figure \ref{fig:maximum_masses}.  Only the amplitude is mildly affected, but the point estimates for the distributions remain the same. 

In the subsequent analysis, we examined the impact of changing the treatment of individual pulsar masses in the ``accuracy-dependent'' model for the four systems mentioned in the previous section (PSR B1957+20, PSR J1311-3430, PSR B1516+02B, PSR J1748-2021B). In order to do that, we kept the treatment described in Section \ref{sec:alsing_model} for all systems, with the exception of those mentioned above, which are now also assumed to follow a normal distribution. The result is shown by the dot-dashed line in the left panel of Figure  \ref{fig:maximum_masses}. By changing the likelihood treatment for only these particular systems, the marginalized distribution of $m_{max}$ shifts to the right, and the probability of higher $m_{max}$ values is increased. The dot-dashed line in the left panel tends to the solid line in the right panel as we assume more systems to be modeled with Gaussians, and equalizes when all systems are treated as Gaussian. As we can note, the lower individual mass estimates obtained when a flat distribution over $i$ is invoked (and discussed in Section \ref{sec:eff_ind_mass}) are reflected, consequently, in a lower estimation of $m_{max}$.

\begin{figure}[H]
    \begin{subfigure}
        \centering
        \includegraphics[width=.4\textwidth]{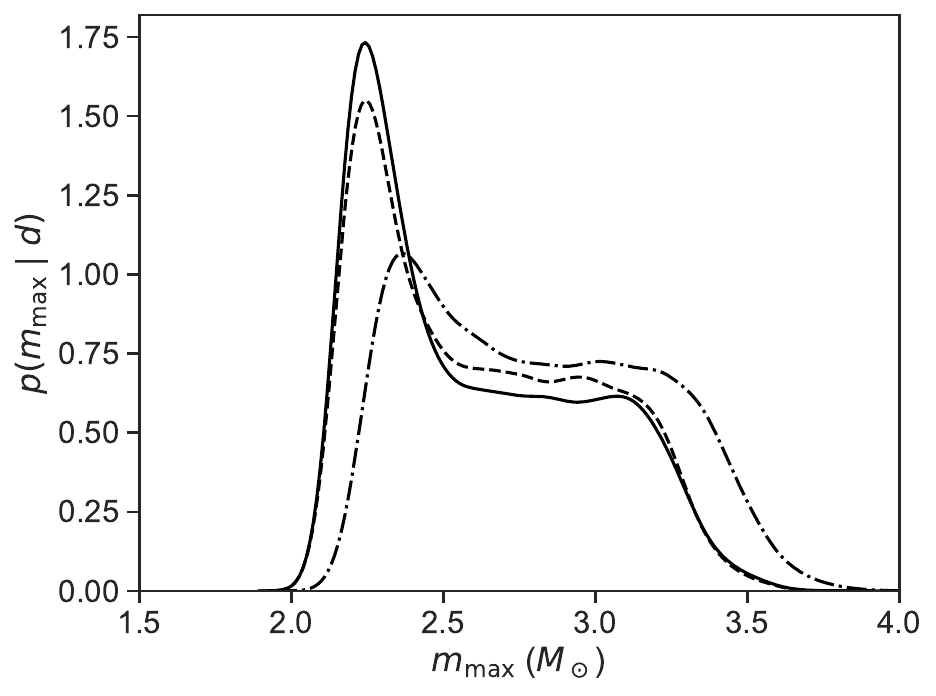}\label{fig:alsingmmax}
    \end{subfigure} %
    \begin{subfigure}
        \centering
        \includegraphics[width=.4\textwidth]{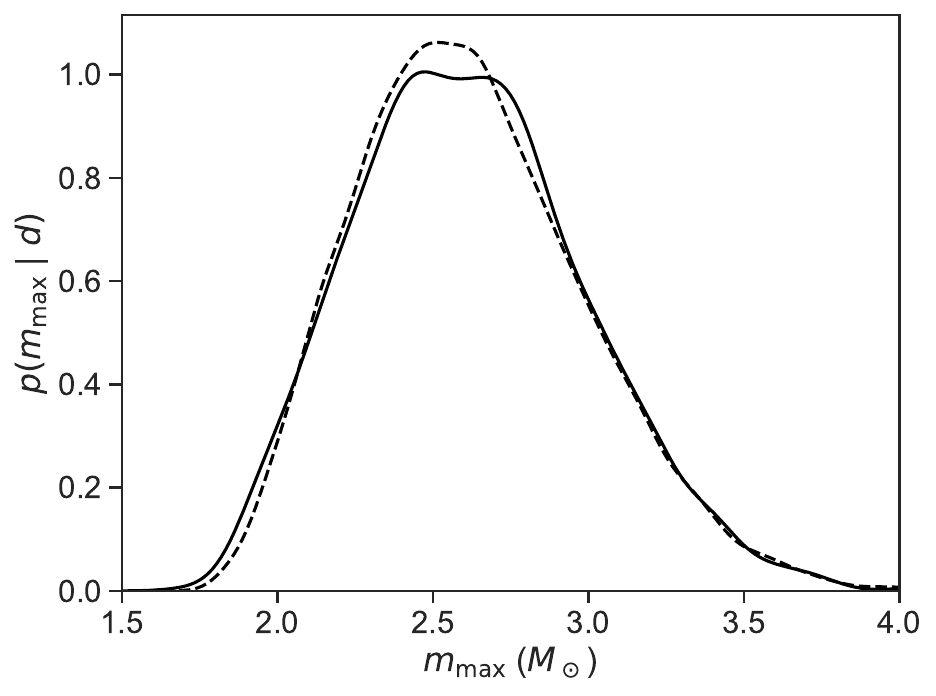}\label{fig:gaussianmmax}
    \end{subfigure} %
\caption{Marginal \textit{posterior} distribution of $m_{max}$ sampled from the ``accuracy-dependent'' model (left panel) and the ``accuracy-independent'' model (right panel). The solid line represents the analysis assuming the individual pulsars to be sampled from 3 types of likelihoods (left panel), and to be sampled from normal distributions only (right panel). In the dashed curve, we analyze the effect of assuming the masses found from $\gamma$-ray observations, listed in Table \ref{tab:gamma_rays}.}\label{fig:maximum_masses}
\end{figure}

As demonstrated, the prior assumption regarding the orbital inclination angle significantly influences the determination of the mass limits, emphasizing the importance of a cautious approach. The high amplitude of values close to $2~M_\odot$ in the left panel of Figure  \ref{fig:maximum_masses} is a consequence of lower estimates of individual masses found in this particular approach, leading to a density accumulation around $2~M_\odot$ and a significant decrease in the probability for values $\geq 2.4~M_\odot$. 

In this scenario, the ``accuracy-independent'' model is preferred, since the individual sampled masses are consistent with {values derived from} observations. The ideal scenario, however, would involve the adoption of the ``accuracy-dependent'' model, with consideration of observational constraints on the inclination angle ($i$) for each binary system. This meticulous treatment remains a subject for a future work.

\subsection{Posterior Predictive Check}
In Bayesian analysis, a crucial step to validate a model is to assess whether predictive simulated data resemble the observed data. In essence, we aim to understand whether new observations drawn from the posterior distribution would be consistent with the existing observed sample. Any significant disparity might indicate a misfit. This analytical process is known as a \textit{posterior predictive check} (PPC). One approach to conducting a PPC is to visually compare summaries of real data with those of simulated data. In addition, it can be valuable to quantitatively assess the level of discrepancy. This can be achieved by defining a ``test quantity'' ($T$), often chosen as the mean, and calculating a Bayesian p-value. This p-value indicates the probability that the test quantity of simulated data ($T^{sim}$) exceeds the observed test \mbox{quantity ($T$):}
\begin{equation}
  p = P(T(m^{sim}) > T(m) | m ).
\end{equation}

Following the detection of GW170817, several studies \cite{ai2020constraints, rezzolla2018, margalit2017, ruiz2018, shibata2019} derived mass limits for NSs based on observational constraints set by the GW event. All these analyses resulted in maximum masses that are significantly lower than the $m_{max}$ we found for the ``accuracy-independent'' model.

Our goal is to investigate the maximum value of the distribution and, faced with the substantial differences we have identified for this quantity between the two models we treated here and with the values derived from GW data, we sought further investigation to determine which value of $m_{max}$ complies with the available sample of NS masses. Consider that we are analyzing a distribution without specifying that it has to be truncated, the question we want to address is: what would be the value consistent with an upper threshold for this distribution?

For this purpose, we conducted a new MCMC sampling from a non-truncated Gaussian mixture, with a normal likelihood for all pulsars. The only difference to the previous model is the absence of the denominator in Equation (\ref{eq:trunc_likelihood}). In Table \ref{tab:sampling}, we summarize the mean and standard deviation of each model parameter. 

\begin{table}[H]
\caption{Summary of marginal \textit{posterior} distribution of each parameter in the non-truncated bimodal model. These results were used to generate 10,000 synthetic bimodal Gaussian distributions.} 
\label{tab:sampling} 
\tabcolsep=.15cm               
\begin{tabular}{c  c  c c c c } 
\toprule           
     \boldmath{  $r_1$} &\boldmath{ $r_2$} &\boldmath{ $\mu_1$ }& \boldmath{$\mu_2$} & \boldmath{$\sigma_1$} &\boldmath{ $\sigma_2$ }\\
 \midrule 
      $0.527 \pm 0.090$ & $0.473 \pm 0.090$ & $1.354 \pm 0.026$   & $1.801 \pm 0.070$ & $0.095 \pm 0.023$ & $0.301 \pm 0.037$ \\
\bottomrule 
\end{tabular}
\end{table}

Using the mean values reported in Table \ref{tab:sampling}, we then generated 10,000 posterior predictive distributions using the software
 \texttt{MATHEMATICA} \cite{Wolfram}. Subsequently,
 we defined the test quantity, $T$, as the number of elements in the observed sample with masses exceeding a specific value, referred to as ``$m_{max}$''. This approach enables us to examine the upper limit for the distribution. The p-value is derived from the number of times $T^{sim} > T$. To illustrate, the top left panel in Figure \ref{fig:posteriorcheck} offers insight into the probability of detecting new objects with $m > 2.09~M_\odot$ in future observations, which is $47.8\%$, a high value. A p-value close to zero is consistent with a \textit{mass limit}, {i.e.}, a low p-value implies a reduced probability of detecting new objects with masses exceeding the associated value. Through this alternative analysis, we confirm that the distribution of galactic NSs supports high $m_{max}$ values, such as the $2.6~M_\odot$ we found from the ``accuracy-independent'' model.

 It is important to note that in this analysis all masses were modeled with Gaussians. An inconsistency between results found through the PPC and through the ``accuracy-independent'' model would also reveal a flaw in the ``accuracy-independent'' model, which is not the case. This analysis is not intended to be conclusive, but rather complementary.

\begin{figure}[H]
    \begin{subfigure}
        \centering
        \includegraphics[width=.26\textwidth]{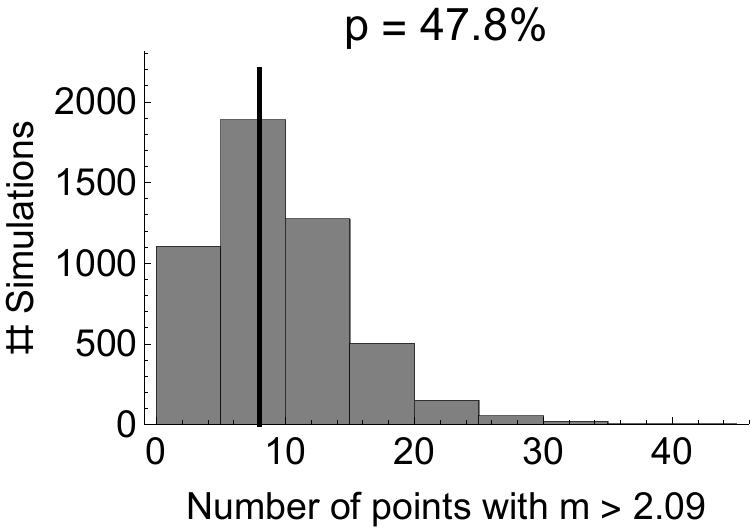}\label{fig:Ai}
    \end{subfigure} %
    \begin{subfigure}
        \centering
        \includegraphics[width=.26\textwidth]{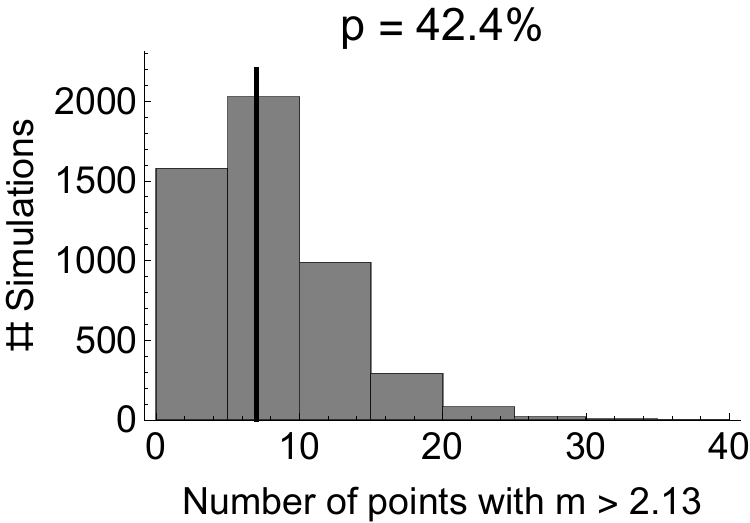}\label{fig:Shao}
    \end{subfigure} %
    \begin{subfigure}
        \centering
        \includegraphics[width=.26\textwidth]{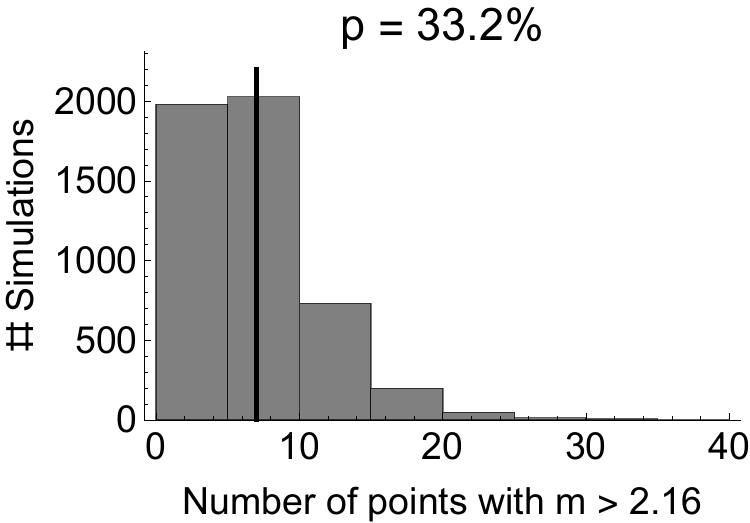}\label{fig:Rezz}
    \end{subfigure} %
    \\[\smallskipamount]
    \begin{subfigure}
        \centering
        \includegraphics[width=.26\textwidth]{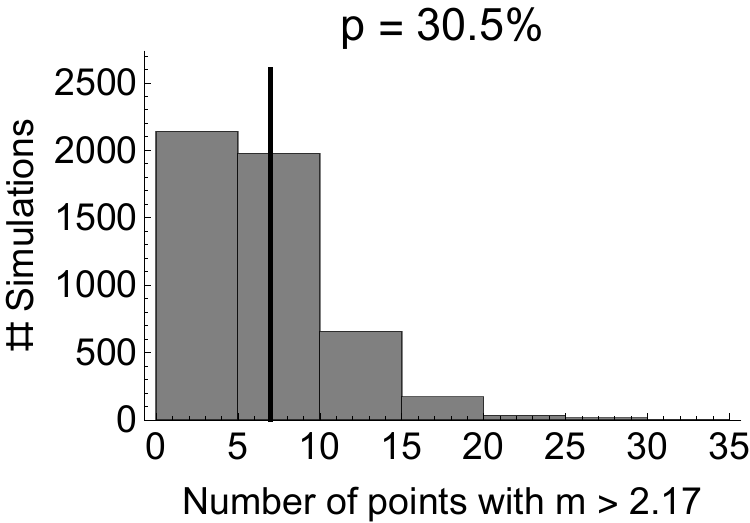}\label{fig:Marg}
    \end{subfigure} %
    \begin{subfigure}
        \centering
        \includegraphics[width=.26\textwidth]{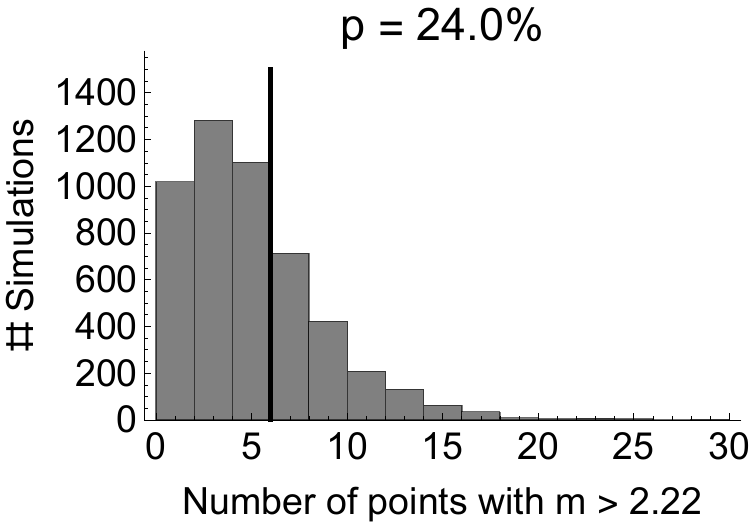}\label{fig:Ruiz}
    \end{subfigure} %
    \begin{subfigure}
        \centering
        \includegraphics[width=.26\textwidth]{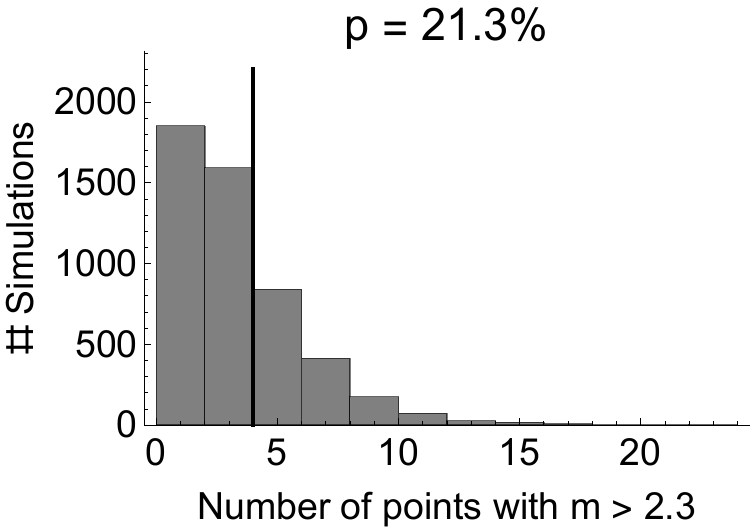}\label{fig:Shibata}
    \end{subfigure} %
    \\[\smallskipamount]\hspace{-5.5cm}
    \begin{subfigure}
        \centering
        \includegraphics[width=.26\textwidth]{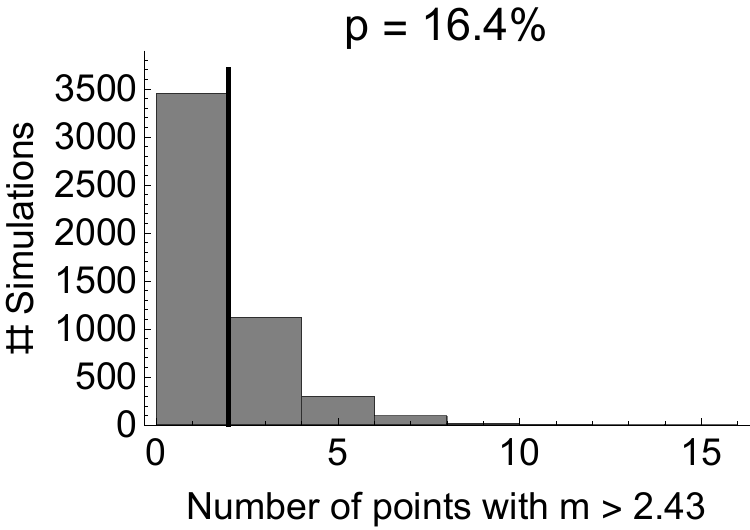}\label{fig:Ai2}
    \end{subfigure} %
    \begin{subfigure}
        \centering
        \includegraphics[width=.26\textwidth]{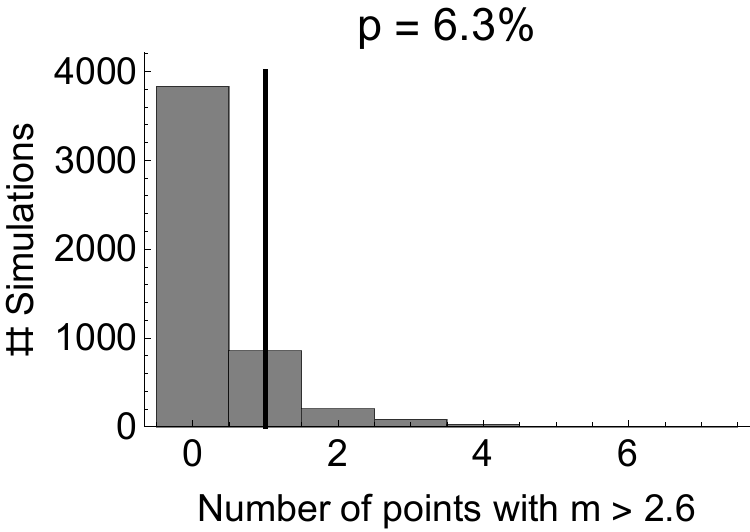}\label{fig:Rocha}
    \end{subfigure} %
\caption{Posterior
 predictive check on the two-Gaussian model without truncation. The purpose is to investigate the tail of distributions. High p-values indicate that values higher than the one specified in the label are very common, thus they cannot be pointed to as valid thresholds. The adopted $m_{max}$ from NS--NS mergers are, from left to right and top to bottom, \citet{ai2020constraints} with $2.09^{+0.11}_{-0.09}~M_\odot$; \citet{shao2020b} with $2.13^{+0.08}_{-0.07}~M_\odot$;  \citet{rezzolla2018} with $2.16^{+0.17}_{-0.15}~M_\odot$; \citet{margalit2017} with $2.17~M_\odot$; \citet{ruiz2018} with 2.16--2.28~$M_\odot$; \citet{shibata2019} with $2.3~M_\odot$; \mbox{\citet{ai2020constraints}} with $2.43^{+0.10}_{-0.08}~M_\odot$. The last panel represents our result for the ``accuracy-independent'' model. We used the mean value of each referenced work, since they cover the whole range of high masses \mbox{reasonably well}.}\label{fig:posteriorcheck}
\end{figure}

\section{Discussions and Conclusions}
The old idea of a unique formation channel for NSs and the existence of a canonical mass have long been invalidated, possibly pointing towards a variety of formation mechanisms, including electron capture supernova and accretion-induced collapses. The current debate revolves around whether the mass limit supported by NSs is above or below the range of 2.2--2.3~$M_\odot$. 

Recent analyses of the galactic sample of NSs have found evidence for a sharp cut-off at $m_{max} < 2.3~M_\odot$ \cite{alsing2018evidence, shao2020a}, even though the number of high-mass estimates in observed NSs continues to grow. The most recent addition to this sample is PSR J0952-0607, with a reported mass of $2.35 \pm 0.17~M_\odot$ \cite{Romani2022heaviest}. On the other hand, observations from the gravitational wave event, GW170817, have also played a crucial role in constraining the mass limit, with results falling in the interval between 2.1 and $2.3~M_{\odot}$ \cite{ai2020constraints, shao2020b, rezzolla2018, margalit2017, ruiz2018, shibata2019}, in line with previous galactic analyses. However, the detection of the GW190814 signal, originating from the merger of a $23~M_\odot$ BH with an unidentified $m = 2.59_{-0.09}^{+0.08}~M_\odot$ companion, raised a new tension. An analysis of the BBH merging population, performed by the LV team \cite{abbott2021population}, found the GW190814 to be an outlier, {i.e.}, it is potentially associated with an NS--BH merger. In addition, a maximum mass constraint for non-spinning NSs was also derived from the available LV sample of NS--BH events, with a result of  $2.7^{+0.5}_{-0.4}~M_{\odot}$ \citep{ye2022inferring}. More recently, an analysis combining as many astronomical observations as possible led to a preferred threshold value of 2.49--2.52~$M_\odot$ \cite{Ai2023constraints}, while \citet{fan2023maximum} placed a minimum of $2.25^{+0.08}_{-0.07}~M_\odot$ to $M_{TOV}$, two completely independent pieces of evidence, to solve the puzzle of the maximum mass value.

To address this ongoing tension, we performed in this work an analysis of the mass distribution of galactic NSs to investigate the impact of the assumption of a uniform distribution between 0 and 1, associated with the orbital inclination angle $i$. As previously mentioned, this parameter is responsible for the largest uncertainties in mass measurements and requires careful consideration. Our findings reveal that this uniform assumption results in individual sampled masses that fall below the constraints set  {from observational modeling} (see Section \ref{sec:eff_ind_mass}). As a consequence, the distribution threshold $m_{max}$ also shows a preference for lower values, at $\sim$2.2~$M_\odot$.

In the ``accuracy-independent'' model, all pulsars are modeled as normal distributions. {The mass values reported in Table \ref{tab:mass} were derived while taking into account observational constraints at $i$, which in turn can be model-dependent.}  {This approach} is consistent with a scenario favoring the existence of extremely massive NSs ($\sim $2.6~$M_\odot$). A complementary analysis (PPC in Section \ref{fig:posteriorcheck}) strengthens the possibility of a high $m_{max}$ for the galactic population, in line with a few conclusions derived from GW analyses, mentioned throughout this work. The discrepancy between the two approaches we treat in this work revealed that it is necessary to be careful when making assumptions over $i${, since different hypotheses over this quantity can lead to drastically different pulsar masses, consequently influencing the $m_{max}$ determination}. An ideal approach would involve implementing the ``accuracy-dependent'' model while accounting for the distribution of $i$ for each particular system, constrained from observational data, as performed for the case of black hole masses \mbox{in \cite{lucas2023galaxies}}, {where each object was analyzed individually and it was checked that the different Keplerian parameter values employed in each mass estimate did not rely on inconsistent assumptions. A similar treatment for NS masses remains a subject for future work.}

Although determining the orbital inclination of a pulsar binary is a challenging task, {we showed in this work that} the constraints obtained from observations are crucial information about the system and must be treated accordingly. Whether the adopted spectrophotometric model adopted to set these constraints is the ideal or not is another question {and needs to be investigated further}. The work of \citet{clark2023neutron}, if confirmed by future research, can help solve the problem of NS mass accuracy, providing a solution to the NS maximum mass puzzle, which in turn has strong consequences for the supranuclear equation of state. But for now, based on the galactic sample, we cannot rule out the possibility that extremely massive NSs ($m > 2.3~M_\odot$) exists in nature.


\vspace{6pt} 



\authorcontributions{Conceptualization, L.S.R. and J.E.H.;
methodology, L.S.R.; software, L.S.R.; validation, L.S.R. and L.M.d.S.;
formal analysis, L.S.R. and L.M.d.S.; data curation, L.S.R.; writing---original draft preparation, L.S.R. and J.E.H.; writing---review and editing, L.G.B., G.Y.C., and M.G.B.d.A. All authors have read and agreed to the published version of \mbox{the manuscript.}}

\funding{This research was funded by FAPESP Agency (S\~ao Paulo State)  grant number 2020/08518-2. L.S.R was funded by Capes Agency (Brazil). L.M.S. received funding from the CNPq (Brazil), grant number 140794/2021-2.}

\institutionalreview{Not applicable.}

\informedconsent{Not applicable.}

\dataavailability{Data are contained within the
> article.}

\conflictsofinterest{The authors declare no conflicts of interest. The funding agencies had no role in the design of the study; in the collection, analyses, or interpretation of data; in the writing of the manuscript; or in the decision to publish the results.} 



\abbreviations{Abbreviations}{
The following abbreviations are used in this manuscript:\\

\noindent 
\begin{tabular}{@{}ll}
NS & Neutron Star\\
GR & General Relativity\\
Fe & Iron \\
TOV & Tolman-Oppenheimer-Volkoff\\
EoS & Equation of state\\
BH & Black Hole\\
O-Ne-Mg & Oxygen-Neon-Magnesium\\
AIC & Accretion-Induced-Collapse\\
WD & White-Dwarf\\
GW & Gravitational Wave\\
LV & LIGO-Virgo\\
BBH & Binary Black Hole\\
ToA & Time of arrival\\
NICER & Neutron star Interior Composition ExploreR\\
pK & post-Keplerian\\
DNS & Double Neutron Star\\
MCMC & Markov Chain Monte Carlo\\
HPDI & Highest posterior density interval\\
MS & Main sequence\\
PPC & Posterior predictive check

\end{tabular}
}

\appendixtitles{yes} 
\appendixstart
\appendix
\section[\appendixname~\thesection]{Sample of Neutron Stars}\label{app:table}
The complete sample of NSs with mass constraints is displayed in the Table below. 

\textls[-25]{In the ``accuracy-dependent'' model, for systems where $f$ and $m_t$ of $q$ values are reported, we have used them to marginalize individual pulsar masses through \mbox{Equations (\ref{eq:total_mass_like}) and (\ref{eq:mass_ratio_like})}.} For the ``accuracy-independent'' model we sampled pulsar's mass with values provided in the 6th column of Table \ref{tab:mass}. The three systems without individual mass constraints were left out from the second analysis. Since they are DNS systems, recognized to have a distribution way below $2~M_\odot$, they do not contribute significantly for the $m_{max}$ inference. Furthermore, for DNS systems without precise constraints on individual masses (from J1018-1523 to J2140-2311B), we considered only the mass of the pulsar component. 


\begin{table}[H] 
    \tablesize{\scriptsize }
\caption{Neutron star measurements for 125 binary systems, with $1\sigma$ uncertainties.}\label{tab:mass} 
\setlength{\cellWidtha}{\textwidth/7-2\tabcolsep+.2in}
\setlength{\cellWidthb}{\textwidth/7-2\tabcolsep-0in}
\setlength{\cellWidthc}{\textwidth/7-2\tabcolsep-0in}
\setlength{\cellWidthd}{\textwidth/7-2\tabcolsep-0in}
\setlength{\cellWidthe}{\textwidth/7-2\tabcolsep-0in}
\setlength{\cellWidthf}{\textwidth/7-2\tabcolsep-0in}
\setlength{\cellWidthg}{\textwidth/7-2\tabcolsep-.2in}
\scalebox{1}[1]{\begin{tabularx}{\textwidth}{>{\raggedright\arraybackslash}m{\cellWidtha}>{\raggedright\arraybackslash}m{\cellWidthb}>{\raggedright\arraybackslash}m{\cellWidthc}>{\raggedright\arraybackslash}m{\cellWidthd}>{\raggedright\arraybackslash}m{\cellWidthe}>{\raggedright\arraybackslash}m{\cellWidthf}>{\raggedright\arraybackslash}m{\cellWidthg}}  

\toprule
{\textbf{Pulsar}} &{\textbf{Type}} &{\boldmath{$f$ [$M_\odot$]}} &{\boldmath{$m_t$ [$M_\odot$]}} &{\boldmath{$q$}} & {\boldmath{$m_p$ [$M_\odot$]}} & 
{\textbf{Reference}}\\
\midrule
%
%
%
2S 0921-630 & x-ray/optical &  &  &  & 1.44 $\pm$ 0.1 & \cite{steeghs2007mass} \\
4U 1538-522 & x-ray/optical &  &  &  & 1.02 $\pm$ 0.17 & \cite{falanga2015ephemeris} \\
4U 1608-52 & x-ray/optical &  &  &  & 1.57 $\pm$ 0.29 & \cite{ozel2016dense} \\
4U 1700-377 & x-ray/optical &  &  &  & 1.96 $\pm$ 0.19 & \cite{falanga2015ephemeris} \\
4U 1702-429 & x-ray/optical &  &  &  & 1.9 $\pm$ 0.3 & \cite{nattila2017} \\
4U 1724-207 & x-ray/optical &  &  &  & 1.81 $\pm$ 0.31 & \cite{ozel2016dense} \\
4U 1820-30 & x-ray/optical &  &  &  & 1.77 $\pm$ 0.27 & \cite{ozel2016dense} \\
4U 1822-371 & x-ray/optical &  &  &  & 1.96 $\pm$ 0.36 & \cite{munoz2005k} \\
Cen X-3 & x-ray/optical &  &  &  & 1.57 $\pm$ 0.16 & \cite{falanga2015ephemeris} \\
Cyg X-2 & x-ray/optical &  &  &  & 1.71 $\pm$ 0.21 & \cite{casares2010mass} \\
EXO 0748-676 & x-ray/optical &  &  &  & 2.01 $\pm$ 0.21 & \cite{knight2022} \\
EXO 1722-363 & x-ray/optical &  &  &  & 1.91 $\pm$ 0.45 & \cite{falanga2015ephemeris} \\
EXO 1745-248 & x-ray/optical &  &  &  & 1.65 $\pm$ 0.26 & \cite{ozel2016dense} \\
Her X-1 & x-ray/optical &  &  &  & 1.073 $\pm$ 0.358 & \cite{rawls2011refined} \\
J013236.7+303228 & x-ray/optical &  &  &  & 2.0 $\pm$ 0.4 & \cite{bhalerao2012neutron} \\
J0212.1+5320 & x-ray/optical &  &  &  & 1.85 $\pm$ 0.29 & \cite{shahbaz2017} \\
J0427.9-6704  & x-ray/optical &  &  &  & 1.86 $\pm$ 0.11 &  \cite{strader2019} \\
J0846.0+2820  & x-ray/optical &  &  &  & 1.96 $\pm$ 0.41 &  \cite{strader2019} \\
J0952-0607 & x-ray/optical &  &  &  & 2.35 $\pm$ 0.17 &  \cite{Romani2022heaviest} \\
J1023+0038 & x-ray/optical &  &  &  & 1.65 $\pm$ 0.16 &  \cite{strader2019} \\
J1048+2339 & x-ray/optical &  &  &  & 1.96 $\pm$ 0.22 &  \cite{strader2019} \\
J1301+0833  & x-ray/optical &  &  &  & 1.60 $\pm$ 0.23 &  \cite{kandel2022optical} \\          
J1311-3430  & x-ray/optical & & & & $2.22\pm0.10$ &  \cite{kandel2022optical} \\       
J1417.7-4407  & x-ray/optical &  &  &  & 1.62 $\pm$ 0.3 &  \cite{strader2019} \\
J1555-2908  & x-ray/optical &  &  &  & 1.67 $\pm$ 0.06 &  \cite{kennedy2022measuring} \\
J1653-0158  & x-ray/Optical &  &  &  & 2.15 $\pm$ 0.16 &  \cite{kandel2022optical}  \\
J1723-2837 & x-ray/optical &  &  &  & $1.22\pm0.23$ & \cite{strader2019} \\
J1810+1744  & x-ray/Optical &  &  &  & 2.13 $\pm$ 0.04 &  \cite{kandel2022optical} \\ %
J2039.6-5618  & x-ray/optical &  &  &  & 2.04 $\pm$ 0.31 &  \cite{strader2019} \\
J2129-0429  & x-ray/optical &  &  &  & 1.74 $\pm$ 0.18 &  \cite{strader2019} \\
J2215+5135 & x-ray/optical &  &  &  & 2.28 $\pm$ 0.10 &  \cite{kandel2020atmospheric} \\
J2339-0533 & x-ray/optical &  &  &  & 1.47 $\pm$ 0.09 &  \cite{kandel2020heated} \\
KS 1731-260 & x-ray/optical &  &  &  & 1.61 $\pm$ 0.37 &  \cite{ozel2016dense} \\
LMC X-4 & x-ray/optical &  &  &  & 1.57 $\pm$ 0.11 &  \cite{falanga2015ephemeris} \\
OAO 1657-415 & x-ray/optical &  &  &  & 1.74 $\pm$ 0.3 &  \cite{falanga2015ephemeris} \\
SAX 1748.9-2021 & x-ray/optical &  &  &  & 1.81 $\pm$ 0.31 &  \cite{ozel2016dense} \\
SAX J1802.7-2017 & x-ray/optical &  &  &  & 1.57 $\pm$ 0.25 &  \cite{falanga2015ephemeris} \\
SMC X-1 & x-ray/optical &  &  &  & 1.21 $\pm$ 0.12 &  \cite{falanga2015ephemeris} \\
Vela X-1 & x-ray/optical &  &  &  & 2.12 $\pm$ 0.16 &  \cite{falanga2015ephemeris} \\
XTE J1855-026 & x-ray/optical &  &  &  & 1.41 $\pm$ 0.24 &  \cite{falanga2015ephemeris} \\
XTE J2123-058 & x-ray/optical &  &  &  & 1.53 $\pm$ 0.36 &  \cite{casares2002vlt} \\
B1957+20 & x-ray/optical &  0.005 &   & 69.2 $\pm$ 0.8 & $2.4\pm0.12$&\cite{van2011evidence} \\    
J1740-5350 & x-ray/optical &  0.002644  &   & 5.85 $\pm$ 0.13 & 1.6$
\pm$0.3 &  \cite{Ferraro2003} \\
J1816+4510 & x-ray/optical &  0.0017607 &   &  9.54 $\pm$ 0.21 & 1.45 $\pm$ 0.38 &  \cite{kaplan2013metal} \\
B1534+12 & NS-NS &  &  &  & 1.3332 $\pm$ 0.0010 &  \cite{fonseca2014comprehensive} \\
B1534+12 Cp & NS-NS &  &  &  & 1.3452 $\pm$ 0.0010 &  \cite{fonseca2014comprehensive} \\
B1913+16 & NS-NS &  &  &  & 1.438 $\pm$ 0.001 &  \cite{weisberg2016relativistic} \\
B1913+16 Cp & NS-NS &  &  &  & 1.390 $\pm$ 0.001 &  \cite{weisberg2016relativistic} \\
B2127+11C & NS-NS &  &  &  & 1.358 $\pm$ 0.010 &  \cite{jacoby2006measurement} \\
B2127+11C Cp & NS-NS &  &  &  & 1.354 $\pm$ 0.010 &  \cite{jacoby2006measurement} \\
J0453+1559 & NS-NS &  &  &  & 1.559 $\pm$ 0.004 &  \cite{martinez2015pulsar} \\
J0453+1559 Cp & NS-NS &  &  &  & 1.174 $\pm$ 0.004 &  \cite{martinez2015pulsar} \\
J0509+3801 & NS-NS &  &  &  & 1.34 $\pm$ 0.08 &  \cite{lynch2012timing} \\
J0509+3801 Cp & NS-NS &  &  &  & 1.46 $\pm$ 0.08 &  \cite{lynch2012timing} \\
J0514-4002A & NS-NS &  &  &  & 1.25 $\pm$ 0.05 &  \cite{ridolfi2019} \\
J0514-4002A Cp & NS-NS &  &  &  & 1.22 $\pm$ 0.05 &  \cite{ridolfi2019} \\
J0737-3039A & NS-NS &  &  &  & 1.338185 $\pm$ 0.000013 &  \cite{Kramer2021} \\
J0737-3039B & NS-NS &  &  &  & 1.248868 $\pm$ 0.000012 &  \cite{Kramer2021} \\
J1756-2251 & NS-NS &  &  &  & 1.341 $\pm$ 0.007 &  \cite{ferdman2014psr} \\ 

J1756-2251 Cp & NS-NS &  &  &  & 1.230 $\pm$ 0.007 &  \cite{ferdman2014psr} \\
J1757-1854 & NS-NS &  &  &  & 1.3406 $\pm$ 0.0005 &  \cite{cameron2022news} \\
J1757-1854 Cp & NS-NS &  &  &  & 1.3922 $\pm$ 0.0005 &  \cite{cameron2022news} \\
J1807-2500B & NS-NS &  &  &  & 1.3655 $\pm$ 0.0021 &  \cite{lynch2012timing} \\
J1807-2500B Cp & NS-NS &  &  &  & 1.2064 $\pm$ 0.0020 &  \cite{lynch2012timing}  \\
J1829+2456 & NS-NS & &  &  & 1.306 $\pm$ 0.007 &  \cite{haniewicz2021precise} \\
J1829+2456 Cp & NS-NS &  & &  & 1.299 $\pm$ 0.007 &  \cite{haniewicz2021precise} \\
 \bottomrule
\end{tabularx}}   \end{table}

\begin{table}[H]\ContinuedFloat
\caption{{\em Cont.}}
\label{tab:mass} 
    \tablesize{\scriptsize }
\setlength{\cellWidtha}{\textwidth/7-2\tabcolsep+.2in}
\setlength{\cellWidthb}{\textwidth/7-2\tabcolsep-0in}
\setlength{\cellWidthc}{\textwidth/7-2\tabcolsep-0in}
\setlength{\cellWidthd}{\textwidth/7-2\tabcolsep-0in}
\setlength{\cellWidthe}{\textwidth/7-2\tabcolsep-0in}
\setlength{\cellWidthf}{\textwidth/7-2\tabcolsep-0in}
\setlength{\cellWidthg}{\textwidth/7-2\tabcolsep-.2in}
\scalebox{1}[1]{\begin{tabularx}{\textwidth}{>{\raggedright\arraybackslash}m{\cellWidtha}>{\raggedright\arraybackslash}m{\cellWidthb}>{\raggedright\arraybackslash}m{\cellWidthc}>{\raggedright\arraybackslash}m{\cellWidthd}>{\raggedright\arraybackslash}m{\cellWidthe}>{\raggedright\arraybackslash}m{\cellWidthf}>{\raggedright\arraybackslash}m{\cellWidthg}}  

\toprule
{\textbf{Pulsar}} &{\textbf{Type}} &{\boldmath{$f$ [$M_\odot$]}} &{\boldmath{$m_t$ [$M_\odot$]}} &{\boldmath{$q$}} & {\boldmath{$m_p$ [$M_\odot$]}} & 
{\textbf{Reference}}\\
\midrule

J1906+0746 & NS-NS &  &  &  & 1.291 $\pm$ 0.011 &  \cite{van2015binary} \\
J1906+0746 Cp & NS-NS &  &  &  & 1.322 $\pm$ 0.011 &  \cite{van2015binary} \\
J1913+1102 & NS-NS &  &  &  & 1.62 $\pm$ 0.03 &  \cite{ferdman2020asymmetric} \\
J1913+1102 Cp & NS-NS &  &  &  & 1.27 $\pm$ 0.03 &  \cite{ferdman2020asymmetric} \\
J1018-1523 & NS-NS & 0.238062 & 2.3 $\pm$ 0.3 &  &  &  \cite{2023ApJ...944..154S} \\
J1325-6253 & NS-NS & 0.1415168 & 2.57 $\pm$ 0.06 &  & 1.37 $\pm$ 0.27 &  \cite{sengar2022high} \\
J1411+2551 & NS-NS & 0.1223898 & 2.538 $\pm$ 0.022 &  &  &  \cite{2017ApJ...851L..29M}\\
J1759+5036 & NS-NS & 0.081768 & 2.62 $\pm$ 0.03 &  & 1.52 $\pm$ 0.26 &  \cite{2021ApJ...922...35A} \\
J1811-1736 & NS-NS & 0.128121 & 2.57 $\pm$ 0.10 &  & 1.34 $\pm$ 0.16 &  \cite{corongiu2007} \\
J1930-1852 & NS-NS & 0.34690765 & 2.54 $\pm$ 0.03 &  &  &  \cite{swiggum2015} \\
J1946+2052 & NS-NS & 0.268184 & 2.50 $\pm$ 0.04 &  & 1.25 $\pm$ 0.15 &  \cite{stovall2018} \\
J2140-2311B & NS-NS & 0.2067 & 2.53 $\pm$ 0.08 &  & 1.3 $\pm$ 0.2 &  \cite{Balakrishnan2023} \\
B1855+09 & NS-WD &  &  &  & 1.54 $\pm$ 0.13 &  \cite{reardon2021parkes} \\
J0337+1715 & NS-WD &  &  &  & 1.4401 $\pm$ 0.0015 &  \cite{Voisin2020} \\
J0348+0432 & NS-WD &  &  &  & 2.01 $\pm$ 0.04 &  \cite{antoniadis2016} \\
J0437-4715 & NS-WD &  &  &  & 1.44 $\pm$ 0.07 &  \cite{reardon2016} \\
J0621+1002 & NS-WD &  &  &  & 1.53 $\pm$ 0.15 &  \cite{Kasian2012} \\
J0740+6620 & NS-WD &  &  &  & 2.08 $\pm$ 0.07 &  \cite{fonseca2021} \\
J0751+1807 & NS-WD &  &  &  & 1.64 $\pm$ 0.15 &  \cite{desvignes2016high} \\
J0955-6150  & NS-WD &  &  &  & 1.71 $\pm$ 0.03 &  \cite{serylak} \\
J1012+5307  & NS-WD &  &  &  & 1.72 $\pm$ 0.16 &  \cite{mata2020} \\
J1017-7156 & NS-WD & & & & $2.0\pm0.8$ & \cite{reardon2021parkes}\\
J1022-1001 & NS-WD & & & & $1.44\pm0.44$ & \cite{reardon2021parkes}\\
J1125-6014  & NS-WD &  &  &  & 1.68 $\pm$ 0.16 &  \cite{Shamohammadi2023} \\
J1141-6545 & NS-WD &  &  &  & 1.27 $\pm$ 0.01 &  \cite{bhat2008gravitational} \\
J1528-3146  & NS-WD &  &  &  & 1.61 $\pm$ 0.14 &  \cite{berthereau2023} \\
J1600-3053 & NS-WD &  &  &  & 2.06 $\pm$ 0.42 &  \cite{reardon2021parkes} \\
J1614-2230 & NS-WD &  &  &  & 1.94 $\pm$ 0.03 &  \cite{Shamohammadi2023} \\
J1713+0747 & NS-WD &  &  &  & 1.28 $\pm$ 0.08 &  \cite{reardon2021parkes} \\
J1738+0333 & NS-WD &  &  &  & 1.47 $\pm$ 0.07 &  \cite{antoniadis2012relativistic} \\
J1741+1351 & NS-WD &  &  &  & 1.14 $\pm$ 0.34 &  \cite{kirichenko2020searching} \\
J1748-2446am & NS-WD &  &  &  & 1.649 $\pm$ 0.074 &  \cite{andersen2018fourier} \\
J1802-2124 & NS-WD &  &  &  & 1.24 $\pm$ 0.11 &  \cite{ferdman2010precise} \\
J1811-2405 & NS-WD &  &  &  & 2.0 $\pm$ 0.65 &  \cite{Ng2020} \\
J1909-3744 & NS-WD &  &  &  & 1.45 $\pm$ 0.03 &  \cite{Shamohammadi2023} \\
J1910-5958A  & NS-WD &  &  &  & 1.55 $\pm$ 0.07 &  \cite{corongiu2023psr} \\
J1918-0642 & NS-WD &  &  &  & 1.29 $\pm$ 0.10 &  \cite{arzoumanian2018nanograv} \\
J1933-6211  & NS-WD &  &  &  & 1.40 $\pm$ 0.25 &  \cite{Geyer2023} \\
J1946+3417 & NS-WD &  &  &  & 1.828 $\pm$ 0.022 &  \cite{barr2017massive} \\
J1949+3106 & NS-WD &  &  &  & 1.34 $\pm$ 0.16 &  \cite{zhu2019} \\
J1950+2414 & NS-WD &  &  &  & 1.496 $\pm$ 0.023 &  \cite{zhu2019} \\
J1959+2048 & NS-WD &  &  &  & 2.18 $\pm$ 0.09 &  \cite{kandel2020atmospheric} \\
J2043+1711 & NS-WD &  &  &  & 1.38 $\pm$ 0.13 &  \cite{arzoumanian2018nanograv} \\
J2045+3633  & NS-WD & & & & 1.251 $\pm$ 0.021 &  \cite{mckee2} \\
J2053+4650 & NS-WD & & & & 1.40 $\pm$ 0.21 &  \cite{berezina2017discovery} \\
J2222-0137 & NS-WD &  &  &  & 1.831 $\pm$ 0.010 &  \cite{guo2021} \\
J2234+0611 & NS-WD &  &  &  & 1.353 $\pm$ 0.016 &  \cite{stovall2019psr} \\
B1516+02B & NS-WD & 0.000646723 & 2.29 $\pm$ 0.17 &  & $2.08\pm0.19$ &  \cite{freire2008a} \\
B1802-07 & NS-WD & 0.00945034 & 1.62 $\pm$ 0.07 &  & $1.26\pm0.13$ &  \cite{thorsett1999neutron} \\
B2303+46 & NS-WD & 0.246261924525 & 2.64 $\pm$ 0.05 &  & $1.38\pm0.08$ &  \cite{thorsett1999neutron} \\
J0024-7204H & NS-WD & 0.001927 & 1.665 $\pm$ 0.007 &  & 1.41 $\pm$ 0.08 &  \cite{Freire2017, freire2003tuc} \\
J1748-2021B & NS-WD & 0.0002266235 & 2.69 $\pm$ 0.071 &  & $2.74\pm0.21$ &  \cite{freire2008b} \\
J1748-2446I & NS-WD & 0.003658 & 2.17 $\pm$ 0.02 &  & $1.91\pm0.06$ &  \cite{kiziltan2013neutron} \\
J1748-2446J & NS-WD & 0.013066 & 2.20 $\pm$ 0.04 &  & $1.79\pm0.06$ &  \cite{kiziltan2013neutron} \\
J1750-37A & NS-WD & 0.0518649 & 1.97 $\pm$ 0.15 &  & $1.26\pm0.37$ &  \cite{freire2008a} \\
J1823-3021G & NS-WD & 0.0123 & 2.65 $\pm$ 0.07 &  & 2.1 $\pm$ 0.2 &  \cite{ridolfi2021} \\
J1824-2452C & NS-WD & 0.006553 & 1.616 $\pm$ 0.007 &  & $1.31\pm0.25$ &  \cite{begin2006search} \\
J0045-7319 & NS-MS &  &  &  & 1.58 $\pm$ 0.34 &  \cite{thorsett1999neutron} \\
J1903+0327 & NS-MS &  &  &  & 1.667 $\pm$ 0.016 &  \cite{arzoumanian2018nanograv} \\
\bottomrule
\end{tabularx}}
\end{table}

\begin{adjustwidth}{-\extralength}{0cm}
\printendnotes[custom] 

\reftitle{References}

\PublishersNote{}
\end{adjustwidth}
\end{document}